\documentstyle[twoside,fleqn,espcrc2,epsf]{article}

\newcommand{\AmS}{{\protect\the\textfont2
  A\kern-.1667em\lower.5ex\hbox{M}\kern-.125emS}}

\def\spose#1{\hbox to 0pt{#1\hss}}
\def\ltapprox{\mathrel{\spose{\lower 3pt\hbox{$\mathchar"218$}}
 \raise 2.0pt\hbox{$\mathchar"13C$}}}
\def\gtapprox{\mathrel{\spose{\lower 3pt\hbox{$\mathchar"218$}}
 \raise 2.0pt\hbox{$\mathchar"13E$}}}
\def\inapprox{\mathrel{\spose{\lower 3pt\hbox{$\mathchar"218$}}
 \raise 2.0pt\hbox{$\mathchar"232$}}}

\newcommand{\be}{\begin{equation}}
\newcommand{\ee}{\end{equation}}

\def\su3{$SU(3)$}

\newcommand{\pl}{{\rm pl}}
\newcommand{\rt}{{\rm rt}}
\newcommand{\pg}{{\rm pg}}

\newcommand{\sig}{{a^2 \sigma}}
\newcommand{\ssig}{{\sqrt{a^2 \sigma}}}
\newcommand{\kstar}{{K^*}}
\newcommand{\mkstar}{{a m_\kstar}}
\newcommand{\mrho}{{a m_\rho}}
\newcommand{\mN}{{a m_N}}

\def\order{{\cal O}}

\title{Quenched $SU(3)$ hadron spectroscopy using improved fermionic and\\
  gauge actions}

\author{Sara~Collins, Robert~G.~Edwards\thanks{Speaker at the conference},
        Urs~M.~Heller, and John~Sloan\\
        SCRI, Florida State University,
        Tallahassee, FL 32306-4052, USA}
\begin{document}

\begin{abstract}

We present results of quenched $SU(3)$ hadron spectroscopy using
$\order(a)$ improved Wilson fermions. The configurations were
generated using an $\order(a^2)$ improved 6-link $SU(3)$ pure gauge
action at $\beta$'s corresponding to lattice spacings of $0.43$,
$0.25$, $0.20$, $0.18$, and $0.15$ fm.  We find evidence that
fermionic scaling violations are consistent with $\order(a^2)$ errors.

\end{abstract}

\maketitle

\section{INTRODUCTION}

The Symanzik action improvement program has been proposed
\cite{Luscher_85,SW_85} as a way to reduce scaling violations in the
approach to the continuum limit from a lattice action.  In another
contribution,
we report on our preliminary investigations into the
nature of scaling violations inherent in improved actions.
In another proceedings in this volume\cite{Sloan_lat95}, we make
detailed comparisons of Wilson and tadpole-improved Clover fermion
actions on configurations with dynamical fermion content.
In this work, we perform $SU(3)$ quenched spectroscopy using a
one-loop tadpole-improved Symanzik pure gauge action and the
tree-level tadpole-improved Clover fermion action. We
measure the hadron spectrum and the string tension at lattice spacings
$0.15$, $0.18$, $0.20$, $0.25$, and $0.43$ fm. The goal of this work
is to measure the lattice spacing dependence of the hadron
spectrum. We find significant scaling violations in $\mkstar/\ssig$
consistent with $\order(a^2)$.

\section{ACTIONS}

Recently, it was shown \cite{Alford_95}
that the tadpole improved Symanzik inspired pure gauge
action \cite{Luscher_85}
is successful in restoring Lorentz invariance in the static
quark potential even at a lattice spacing of $0.43$fm.
Inspired by this result, we used the action
\begin{displaymath}
S[U] = \beta_\pl \big[\sum_\pl \frac{1}{3}\mbox{\,Re\,Tr\,}U_\pl
+ c_\rt \sum_\rt \frac{1}{3}\mbox{\,Re\,Tr\,}U_\rt
\end{displaymath}
\begin{equation}
\quad + c_\pg \sum_\pg \frac{1}{3}\mbox{\,Re\,Tr\,}U_\pg\big] \label{G_action},
\end{equation}
where
\begin{eqnarray}
c_\rt &=& - {\left(1 + 0.4805\alpha_s\right)}\  /\  {20 u_0^2} \\
c_\pg &=& - {0.03325\alpha_s}\  /\  {u_0^2} \quad . \label{glue_coeff}
\end{eqnarray}
The sums are over plaquettes ($\pl$), $1\times 2$ rectangles
($\rt$), and the six length path $(\mu,\nu,\rho,-\mu,-\nu,-\rho)$ with
$\mu,\nu,\rho$ all different ($\pg$).
We use the measured expectation value of the plaquette to determine
$u_0$. The coupling constant $\alpha_s$ is calculated using the tree
level action,
\begin{eqnarray}
u_0 &=& \left({\frac{1}{3}}\mbox{\,Re\,Tr\,}\langle
  U_\pl\rangle\right)^{1/4} \label{u0} \\
\alpha_s &=& -\frac{\log\left(\frac{1}{3}\mbox{\,Re\,Tr\,}
  \langle U_\pl\rangle\right)} {3.06839} \label{alpha_s}
\end{eqnarray}

\begin{figure}
\vspace*{-4mm} \hspace*{-0cm}
\begin{center}
\epsfxsize = 0.45\textwidth
\leavevmode\epsffile{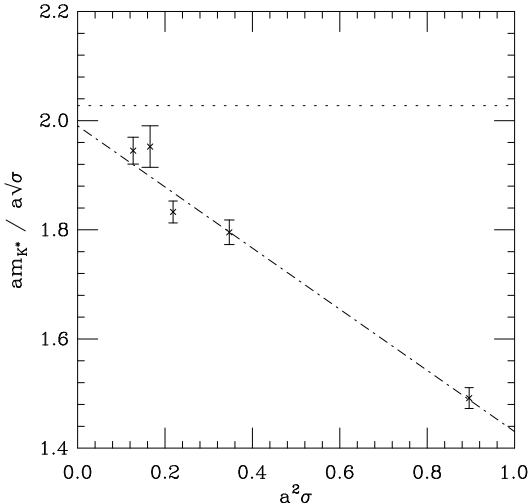}
\end{center}
\vspace*{-2cm}
\caption{
  Scaling plot of $\mkstar$ versus the string tension $\sig$.
  The horizontal line is the physical value using $\sig=440\rm{MeV}$.
  The diagonal line is a least squares fit.
}
\label{kstar_sig}
\end{figure}

Classically, this action has $\order(a^4)$ errors but quantum effects
induce $\order(\alpha^2 a^2)$ errors.  However, it was shown
\cite{Alford_95} that the
action (\ref{G_action}) is insensitive to nonperturbative tuning of the
coefficients suggesting that the $\order(\alpha^2 a^2)$ errors
are small.

To reduce fermionic scaling errors, we use the
tadpole-improved version of the tree-level Clover fermionic
action proposed in \cite{SW_85}.
\begin{eqnarray}
\lefteqn{S_F^{SW} = S_F^W} \nonumber \\
 &-& i c(g_0^2) {\frac{\kappa}{2 u_0^3}}\sum_{x,\pl}
   \overline\psi(x) \sigma_\pl F_\pl(x) \psi(x), \label{F_action}
\end{eqnarray}
where $S_F^W$ is the usual Wilson fermionic action
and $F_\pl$ is a lattice definition of the field strength
tensor.
For the tree level action, $c(g_0^2) = 1$.

With this tree-level action, one expects quantum errors of
$\order(\alpha a)$. However, Naik \cite{Naik_93} computed the one loop
and the dominant two loop contributions to $c(g_0^2)$ and finds that
after tadpole improvement the coefficient of $g_0^2$ is $0.016$.
Therefore, we expect to see only the dominant $\order(a^2)$ fermionic
errors.

\begin{figure}
\vspace*{-4mm} \hspace*{-0cm}
\begin{center}
\epsfxsize = 0.45\textwidth
\leavevmode\epsffile{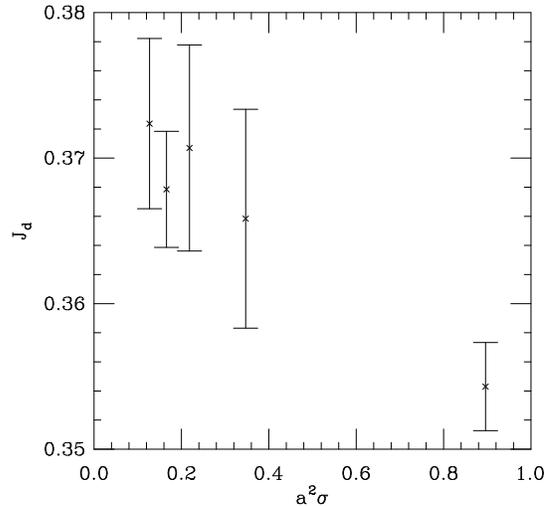}
\end{center}
\vspace*{-2cm}
\caption{
  Scaling plot of $J_d$ versus the string tension $\sig$.
  The physical value is $0.499$.
}
\label{Jdis_fig}
\end{figure}

\begin{figure}
\vspace*{-4mm} \hspace*{-0cm}
\begin{center}
\epsfxsize = 0.45\textwidth
\leavevmode\epsffile{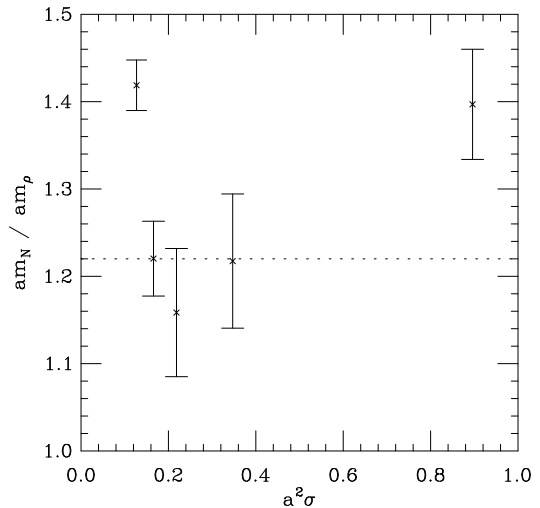}
\end{center}
\vspace*{-2cm}
\caption{
  Scaling plot of the nucleon versus the rho as a function of the
  string tension $\sig$.
  The horizontal line is the physical value of $1.22$.
}
\label{N_rho_sig}
\end{figure}

\section{OBSERVABLES}
To set the scale, $a$, we used the string tension.
We computed finite $T$ approximations to the static quark potential
using time-like Wilson loops $W(\vec R, T)$ which were constructed using
`APE'-smeared spatial links \cite{APE_87}.
On and off-axis spatial paths were used with distances $R=n,
\sqrt{2n}, \sqrt{3n}$ and $\sqrt{5n}$, where $n$ is a positive integer.
The ``effective'' potentials
were fitted using a correlated $\chi^2$ procedure with the spatial
covariance estimated using bootstrap. The string tension was estimated
using the form
\be
V(\vec R) = V_0 + \sigma R - \frac{e}{R} - f\left(G_L(\vec R)
 -  \frac{1}{R}\right).
\ee
The last term takes account of the lattice artifacts at short
distance.  Here, $G_L(\vec R)$ denotes the lattice Coulomb potential
for the Wilson gluonic propagator. We did not have available the
lattice Coulomb potential for the tree-level Symanzik action; however,
we found $G_L(\vec R)$ enabled fits to smaller $R$.

For hadron measurements, we used correlated multi-state fits to
multiple correlation functions as discussed in \cite{Sloan_lat95}. We
computed quark propagators in Coulomb gauge at four
$\kappa$ values for each $\beta$. We used two gaussian
source smearing functions with smeared and local
sinks.  The widths were chosen to have a positive and negative
overlap, respectively, with the first excited state. This improved the
signal for the excited state in a multi-state (``vector'') fit.
We also measured mixed valence content mesons and baryons.

As discussed in \cite{Sloan_lat95}, we tried to avoid human intervention
as much as possible in the fitting procedure. For the multistate fits,
we chose the plateau with $\max{N_{dof} Q \prod_i{\delta M_i/M_i}}$ where
the product is over the relative errors of the ground and first
excited state masses, and $Q$ is the confidence level for a $N_{dof}$
degree of freedom fit.

Once a fitting range was chosen, a bootstrap ensemble of masses was
generated and used for correlated $\chi^2$ chiral fits. We fit the
particle mass according to the ans\"atze
\be
M_{\rm particle} = \left\{
  \begin{array}{l}
     C_0 \ + \ C_2 M_{PS}^2  \\
     C_0 \ + \ C_2 M_{PS}^2 \ + \ C_3 M_{PS}^3  \quad .
  \end{array}
\right.
\label{chiral_ansatz}
\ee
This ansatz was motivated by one loop chiral perturbation corrections
to baryon masses.
We note that recently one loop $M_{PS}^3$ corrections to the vector
mass were derived in \cite{Jenkins_95}.

An important method used to increase statistics is via `$Z(3)$'
fermion sources \cite{Butler_94}. Within the source time-slice source for
quark propagators, smearing functions are placed at regular intervals
along each spatial dimension. These smearing functions are weighted
with random elements chosen according to a $Z(3)$ distribution.  Since
the elements are not correlated, each origin in a mesonic or baryonic
construction contribute independently to the correlation functions. The
distance between origins should be chosen larger than the lattice size
of a physical state or the smearing width used.

\begin{table*}
\begin{center}
\small
\caption{String tension and best fits of masses extrapolated to
physical ratios.}
\label{fit_table}
\begin{tabular}{|c|r|r|c|l|l|l|l|l|}
\hline
$\beta$ & $L$ & $N_{meas}$& $u_0$ & $\sig$ & $\mrho$ &
  $\mkstar$ & $J_d$ & $\mN$ \\
\hline
$7.90$ & 16 & 100 & 0.8848 & 0.1272( 16) & 0.6206( 58) & 0.6937(44) &
  0.3724(58) & 0.881( 1) \\
$7.75$ & 16 & 100 & 0.8800 & 0.1658( 50) & 0.7123( 46) & 0.7950(35) &
  0.3679(40) & 0.869(25) \\
$7.60$ & 16 &  32 & 0.8736 & 0.2184( 13) & 0.7666( 85) & 0.8565(68) &
  0.3707(71) & 0.888(46) \\
$7.40$ &  8 &  68 & 0.8629 & 0.3470( 29) & 0.9482(115) & 1.0577(89) &
  0.3658(75) & 1.154(59) \\
$6.80$ & 16 &  72 & 0.8261 & 0.8961(177) & 1.2706( 71) & 1.4120(41) &
  0.3543(30) & 1.778(70) \\
\hline
\end{tabular}
\end{center}
\end{table*}

\section{SIMULATIONS}

We generated quenched configurations using the action (\ref{G_action})
at $\beta = 7.90$, $7.75$, $7.60$, $7.40$, and $6.80$.
Unfortunately, the non-local nature of the action precluded an
efficient implementation of an over-relaxed / heat-bath algorithm for
the CM-2 at SCRI -- we would need to divide the lattice into 32 sublattices.
We used a Hybrid Monte Carlo algorithm with a second order integration
scheme, step size $\delta=0.1$, and a global acceptance test. We
typically found $> 0.75\%$ acceptance rates.

We determined meson and baryon masses using the
action (\ref{F_action}).
The pseudoscalar masses were tuned to the same physical mass
(in units of $\ssig$) between the various $\beta$.
The vector to pseudoscalar mass ratios varied from about $2.0$ down to
$1.2$.
At the coarsest lattice,
we found a large number of exceptional configurations on an
$8^3\times16$ lattice.
To avoid problems, we used a $16^3\times32$ lattice throughout except
at $\beta = 7.40$. We expect finite volume effects to be small since
the box size is $2.40$fm at $\beta=7.90$.

The results of the
chiral extrapolations of our
best fits are listed in Table~\ref{fit_table} along with the string
tension. Note the lattices are too
coarse to reliably use Sommer's force method since the
lattice spacing is close to the empirical $0.5$fm scale. The lattice
spacings we quote used $\ssig = 440{\rm MeV}$.
For the vector mesons, we used a quadratic fit ansatz of
(\ref{chiral_ansatz}) for $\beta = 7.90$, $7.75$, and $7.60$ using all
$10$ mixed valence states. At $\beta=7.40$, we dropped the two largest
masses and used a for a quadratic fit.

At $\beta=6.80$, we found some interesting results.
First, the assumption of using mixed valence states as distinct
measurements appeared to break down -- the meson masses fluctuated
significantly enough to lower the quality of the fit. For this
reason, we only used the four diagonal masses (same valence content)
for fits.  Also, the smallest pseudoscalar measured had $am_{PS} =
0.9939(13)$. At these large masses, we expect significant deviations
from the continuum dispersion relations. Quite significantly, a
quadratic fit ansatz was clearly violated, and we used a cubic fit
ansatz of (\ref{chiral_ansatz}) with
$N_{dof}=1$. This indicates that scaling violations inherent with large
$a m_q$ are roughly approximated by higher order terms in the chiral
ansatz of (\ref{chiral_ansatz}).

In our chiral fits for the nucleon, we only used the four diagonal masses.
For all propagator fits, we used a two exponential fit with two
smearings.
At $\beta = 7.90$ -- $7.40$, we found that the quadratic ansatz broke
down at the largest mass. However, there was not sufficient signal for
a cubic term. At $\beta=6.80$, we used a cubic ansatz for all four masses.
At $\beta=7.90$, the extrapolated nucleon value is way off scaling. We
think this comes from a lack of statistics due to autocorrelations,
and the error bar is underestimated.

\section{SCALING VIOLATIONS}

Our main result for the determination of scaling violations is in
Figure~\ref{kstar_sig}. We interpolate (or extrapolate) the vector
meson mass to the $K^* / K$ ratio and find that there are significant
scaling violations in the $\mkstar/\ssig$ ratio. Fitting to a line, we
find $Q=0.1$ and the intercept of $1.99(2)$ which is close to the
physical value of $2.03$ using $\ssig=440\rm{MeV}$. This implies that
the assumption of $\order(a^2)$ errors is significant. If we fit to
$\ssig$ assuming $\order(a)$ errors, we find that $Q=0.17$ but the
intercept is $2.22(3)$. This value would imply $\ssig=347{\rm MeV}$
which is probably too low. Indeed, constraining the $\order(a)$ fit to
pass through the value $2.0275$ has $Q=10^{-9}$.
We thus feel reasonably confident that the fermionic scaling errors
are (at least) consistent with $\order(a^2)$.

Following \cite{Lacock_95}, we plot the ratio
\be
J_d = \mkstar \frac{\mkstar - \mrho}{(am_K)^2 - (am_\pi)^2}
  \label{Jdis}
\ee
versus $\sig$ in Figure~\ref{Jdis_fig}. We see that $J_d$ is
reasonably scaling out to $\beta=7.40$ indicating that scaling violations
in the individual masses are nearly canceling in the ratio.
The value of $J_d$ is consistent with other quenched calculations.

In Figure~\ref{N_rho_sig}, we plot the ratio of $\mN / \mrho$. We see
reasonable scaling out to $\beta=7.40$, but the value at $\beta=7.90$
is way off as described above.

The computations done on the CM-2 at S.C.R.I.
This research was supported by DOE contracts
DE-FG05-85ER250000 and DE-FG05-92ER40742.

\end{document}